\documentstyle[twocolumn,prc,aps,epsf,epsfig]{revtex}
 \begin{document}
 
\newcommand{\be}{\begin{eqnarray}}
\newcommand{\ee}{\end{eqnarray}}
\twocolumn[\hsize\textwidth\columnwidth\hsize\csname@twocolumnfalse\endcsname
\title{
Interactions between hadrons are strongly modified
 near the QCD (tri)critical point
}

\author{  E.V. Shuryak  }

\address{
Department of Physics and Astronomy, State University of New York, 
Stony Brook NY 11794-3800, USA
}

\date{\today}
\maketitle

\begin{abstract}
The QCD (tri)critical point, a genuine second-order phase transition,
implies existence of a massless scalar mode, and singular behavior
near it. In this work we however focus on the finite region around it, 
defined by a condition $m_\sigma=2m_\pi$.
We point out that in this region the inter-hadron interaction
should be
dramatically changed. Light sigma should
increase attractive mean field potentials for baryons
and non-Goldstone mesons. 
The same effect can be observed in additional downward shift of the mass of 
vector mesons
$\rho,\omega,\phi$,
 accessible via dilepton experiments.
For pions we predict that the mean change due to light sigma
is in fact a
 $repulsive$ mean potential. The implications of these effects for
collective pion and nucleon flows, radial directed and elliptic,  
are estimated.
Finally, we speculate tha unusual behavior of flows observed by NA49
at 40 GeV PbPb collisions at SPS may be explained by location of the
critical point region nearby.
\end{abstract}
\vspace{0.1in}
]
\begin{narrowtext}
\newpage

\section{Introduction}
\subsection{The QCD (tri)critical point}
   Let me start by briefly reminding well known facts.
Ignoring strangeness and starting with massless u,d quark
theory, one arrives at the situation in which the phase transition
at small baryon chemical potential $\mu$ is a second order
line, changing to a first order line at the so called
   {\em tricritical point}. In this theory it is convenient to
look at pions and sigma as one 4-component field $\phi_i$ and
write Landau-Ginzburg potential in terms of its square
\be \Omega={a\over 2} (\phi_i\phi_i)+ {b\over 4} (\phi_i\phi_i)^2+{c\over 6} (\phi_i\phi_i)^6\ee
The coefficients $a(T,mu),b(T,mu),c(T,mu)$ are some functions. The
 second order line is defined by zero mass condition $a(T,mu)=0$:
at it all 4 fields $\pi,\sigma$ are massless, so the transition belongs
to the universality class of the O(4) spin model. At the tricritical
point additionally the second equation holds
$b(T,mu)=0$ which together with the first one select one critical point.
Here the indices are as in the mean field theory.

In the real world the nonzero quark mass adds term $m\sigma$ (and
other
odd terms) to this nice symmetric Lagrangian and makes it a bit more
compacted. Still straightforward manipulations with this function
tell us how all quantities change along the critical line, see the
second paper \cite{SRS}.
 It makes all fields massive in the vacuum, and
transform the second order line 
 to just a crossover (see Fig.\ref{fig_crit_point_ovals} ).
ri The tricritical point is changed
to just a critical point with Ising exponents and just one massless
mode, $m_\sigma=0$.   The pions are massive there, and $m_\pi^2\sim m^{4/5}$
which is only slightly differs from $m^1$ in G-Oaks-Renner
relation in the vacuum. Below we will ignore small correction to it,
$\sim (\Lambda_{QCD}/m)^{1/10}$,
and assume the pion mass at the critical point is the same as in vacuum.
Obviously, there should be a finite region (the shade area in
Fig.\ref{fig_crit_point_ovals} to be refered to as ``oval'' below) in
which $m_\sigma<2m_\pi$. One consequence of this condition is that
$\sigma\rightarrow 2\pi$ decay is impossible: but the boundary of the oval
will be important for other reasons as well, see section \ref{sec_pipi}.

   Stephanov, Rajagopal and myself \cite{SRS} were the first to
 propose experimental search for the QCD (t)critical point in  heavy ion
collisions, 
by varying the energy of and looking for ``non-monotonic'' signals.
The first specific signal  proposed in that paper was an increased 
{\em event-by-event fluctuations} near it, reminiscent of critical
opalescence well known near other 2-nd order phase transitions. 

Although well motivated, the fluctuations 
 is a rather subtle signal, because (as in fact emphasized in
original papers on event-by-event fluctuations
\cite{Stodolsky,Shu_fluct}) the experimentally observed fluctuations are
most likely to be just equilibrium thermodynamical
fluctuations at the thermal freeze-out.
Therefore, critical fluctuations
can only be seen due to some non-equilibrium ``memory effects''.
Asakawa, Heinz and Muller\cite{AHM} and Jeon and
Koch\cite{koch2}  suggested that large-scale fluctuations of conserved
currents
(such as electric and baryon charges) are more slow to dissipate, and
argued that those may show
the ``primordial QGP''  values, rather than those for hadronic gas at
freezeout. Unfortunately further studies
(e.g. by Stephanov and myself \cite{SS_longfluct}) have shown that the
diffusion from the boundary of experimentally covered begins to
be large enough to wipe it out. Experimentally, at any collision
 energy studied 
so far, the
charge fluctuations seen are close to resonance gas values, while 
small expected ``QGP values'' unfortunately are not observed. 

The second proposal of \cite{SRS} was the so called ``focusing
effects'', a particular deformation of adiabatic cooling lines
in the vicinity of the critical point. Nonaka and Asakawa
\cite{Nonaka} have
provided  a  description of this
effect in significant detail.

 In this paper I would like to reconsider the issue and suggest
(hopefully) more robust signals of the critical point, which would 
not rely on subtle memory effects. One of those is collective flows,
which show the effect {\em accumulated during the whole evolution}
along the cooling path, rather than just its end the freezeout. 
The second, even more direct but more challenging
observable, is the invariant mass spectrum of
dileptons, some of which  produced right when the system's cooling
path
passes the oval, perhaps even
 near the critical point.

\begin{figure}[t!]
\centering
\includegraphics[width=8cm]{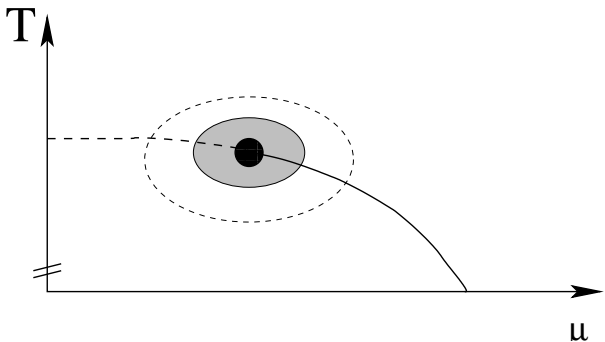}
\vskip 0.4cm
\caption{\label{fig_crit_point_ovals}
Schematic view of the critical point, for 2 flavor QCD with nonzero
quark masses. The black dot mark the position of the critical point,
separating a crossover (the dashed line) from the first order
transition (solid line). 
The shaded oval shows the region where  $m_\sigma<2m_\pi$ and its decay
into pions is forbidden, while the wider ellipse corresponds to the
region in which it is less than twice the average pion energy,$m_\sigma<2<E_\pi(T)$ .
}
\end{figure}

 The reduction in sigma mass near the whole critical
line, and especially at the beginnig
of the crossover region $\mu=0,T=T_c$ were discussed in numerous
models since 1960's, see e.g.  \cite{hepph0204163} 
and references to earlier works  therein. But susceptibilities
obtained in lattice simulations before, at finite $T$ and $zero$ $\mu$, do not
show any large peaks corresponding to light sigma\footnote{
For the quark mass value used in the
calculations, of course.}   . At the critical point, however,
the sigma  mass $must$ be zero by definition, regardless of the
quark masses\footnote{Note that although we know from experiment
how experimental sigma resonance is seen in pion and kaon scattering,
we do not know that for the critical mode. Thus its interaction
with strange quarks may be different from that with light ones:
the corresponding mixing angle between SU(3) singlet and octet is
unknown parameter.
}. Furthermore, this work was
triggered by new lattice data (see next subsection) have now indicated
a quite different behavior of these susceptibilities,
consistent with quite small sigma masses and consequently large
effects at $\mu\approx T_c$.

The discussion of critical properties themselves reduces to the
statement
that it belongs to the simplest Ising universality class, with
standard
critical exponents. The issue of dynamical exponents
and universality class was discussed  by Son and
Stephanov
\cite{SS_modes}: the only relevant conclusion the reader should
be reminded now is that the dynamical index $z$, relating the
available time the system spends near the critical point
\be \tau\sim \xi^z\sim m_\sigma^{-z}\ee
with the maximal correlation length $\xi$, is $z\approx 3$. Therefore, in a
realistic central 
collisions of the heaviest nuclei available, the time limits
the correlation length and the sigma mass by about
\be \xi< 3 \, fm, m_\sigma> 70 \, MeV\ee
Still, a significant reduction of the sigma mass from its vacuum value,
 by about an order of magnitude, may be possible.
As we argue in this work, even a decrease by about factor 2,
bringing us close to ``oval'' in Fig.1,
should result in quite dramatic changes in inter-particle interaction.

\subsection{Recent lattice results}
Discussions of the possible search for the critical point
in experiment has resulted in a search for it on the lattice as well.
Of course this is a very difficult task, as direct simulations at
finite
$\mu$ are impossible due to the notorious sign problem. Two methods
pursued are re-weighting along the critical line \cite{FK} and Taylor
expansion
in powers $\mu/T$. (We remind the reader that $\mu$ in this work is
defined
per quark, so  $\mu/T=1$ means the usual baryon number chemical
potential to be about 500-600 MeV.)

Let me just focus on one most resent study
of the latter type, by Bielefeld-Swansee group \cite{BS_mu6} which
followed such an expansion for the 2-flavor QCD up to terms $O\large((\mu/T)^6\large)$.
The authors conclude that they don't actually see any direct
signatures of the critical point yet, nor in fact any other indications 
for the presence of the first order
transition at larger $\mu$. In the 
covered domain, namely $\mu/T<1$, the Taylor
series  seem to converge
reasonably well.
 This is not surprising, since their quark mass is still
quite large compared to physical one, which is suppose to move
the critical point toward larger $\mu$.
And yet,
a closer look at their results show that 
their numerical data are quite remarkable, they show
surprisingly abrupt changes in the system, and some indications are
perhaps relevant for the issues discussed
in this work.

\begin{figure}[tb]
\begin{center}
\begin{minipage}[c][7.8cm][c]{7.4cm}
\begin{center}
\epsfig{file=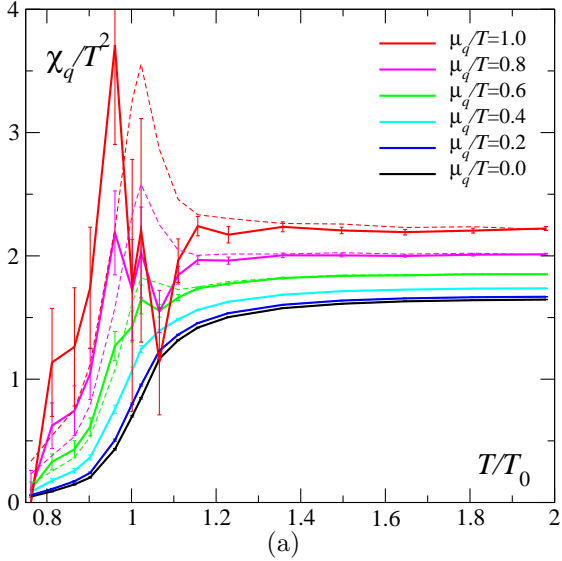, width=7.4cm}\\[-1mm]
(a)
\end{center}
\end{minipage}
\begin{minipage}[c][7.8cm][c]{7.4cm}
\begin{center}
\epsfig{file=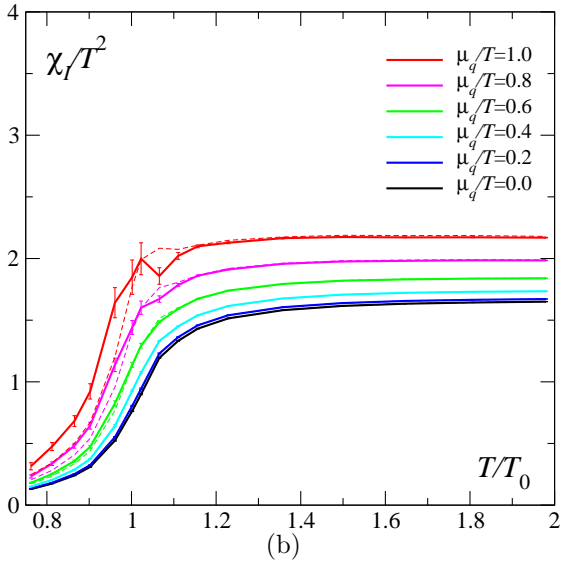, width=7.4cm}\\[-1mm]
(b)
\end{center}
\end{minipage}
\caption{The quark number susceptibility $\chi_q/T^2$ (left) and 
isovector susceptibility $\chi_I/T^2$ (right) as functions of $T/T_0$ for
various $\mu_q/T$ ranging from $\mu_q/T=0$ (lowest curve) rising in
steps of 0.2 to  $\mu_q/T=1$, calculated
from a Taylor series in $6^{th}$ order. Also shown as dashed lines
are results from a $4^{th}$ order expansion in $\mu_q/T$.
}
\label{fig:chiIq}
\end{center}
\end{figure}
In Fig.\ref{fig:chiIq} borrowed from this work we show the 
normalized  quark number susceptibility $\chi_q/T^2$, for the
usual (baryon number) chemical potential, as well as for
isospin-related one. One can see
that, as $\mu/T$ grows, the former one develops a sharp peak 
below $T_c$, while the
 isovector susceptibility does not. 

Is this peak due to 4-th order 2-loop diagram with a sigma meson
exchange, shown in Fig.\ref{fig_diag_chi}(a)? It would show a peak in
$\mu^4$
coefficient (as indeed observed on the lattice) and no peak in the
 isovector case (again, as observed)?

Redlich and Karsch \cite{RK_inprogress} argued that (at least
the left-hand side of the peak $T<T_{peak}$)
 is not due to this diagram, which is
suppressed by a need to excite a baryon in each loop and is thus
suppressed by extra $\sim exp(-M_N/T)\approx 1/300$.
They find it hard to think that even enhancement due to small
sigma mass $\sim 1/m_\sigma^2$ of the diagram (a) can beat that.
 They argue instead that the l.h.s. of the peaks is
quite well described by a baryonic resonance gas\footnote{With masses of
  mesons and baryons correctly adjusted to quark masses used in the
  lattice simulations.} and the usual one-loop diagram (b).

If so, why does the signal drop so rapidly at the right-hand side of
the peak? One possible answer is  {\em rapid melting of 
many baryonic states} above the critical line. Another possibility
is that the susceptibility is 
decreasing because (in spite of deconfinement) the baryon mass
is larger than in vacuum. In fact, the
effective quark
mass (defined as half potential between $\bar q q$ at large separation)
is indeed so large above the critical line, that quark contribution $\sim
exp(-M_q^{eff}(T)/T)$ can get even more suppressed than
that of  the nucleon.

\begin{figure}[t]
\centering
\includegraphics[width=8cm]{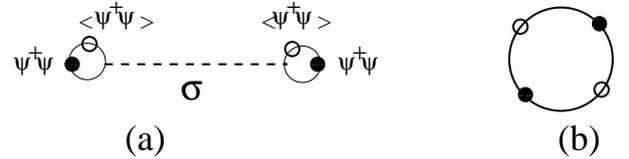}
\vskip 0.2cm
\caption{\label{fig_diag_chi}
Two diagrams responsible for 4-th order part of the density-density
susceptibility.
The black dots stand for the original vector density
 operators,  the open circles are 
for  insertions of the mean vector  density of the matter.
The solid lines are states with the baryon number, 
the baryons below the critical $T$ and quarks/diquarks above it.
The dashed line represents the $\sigma$ meson.
}
\end{figure}

Now let us look at the behavior of the chiral condensate and chiral
susceptibility $\chi_{\bar{\psi} \psi}$, shown in Fig.\ref{fig:ccn}
from the same lattice study. The condensate shows little change, just slight
shift downward in the critical $T$ as $\mu$ grows, as expected.
The susceptibility however  seem to have a larger peak:
in contrast to plots shown above, now there is no
need for extra vector current insertions, so it probably is the
(unsuppressed) sigma contribution. Thus, in spite of larger
uncertainties and similarity to the plot above, I still think this
peak does indeed contain a significant part of a sigma exchange, and
therefore
the peak is the first manifestation of reduced sigma mass.
If the absolute hight of the peak is a measure of  $
1/m_\sigma^2(\mu)$,
one may thing the reduction of the mass itself is by a factor 2 or
so. 
Thus, optimistically, one may think that at least the edge of the
``oval''
of Fig.1 is more or less reached by those simulations.
Needless to say, more work is need to see if this is correct: the
simplest of those is to see a correlator rather than integrated susceptibility.
\begin{figure}
\begin{center}
\epsfig{file=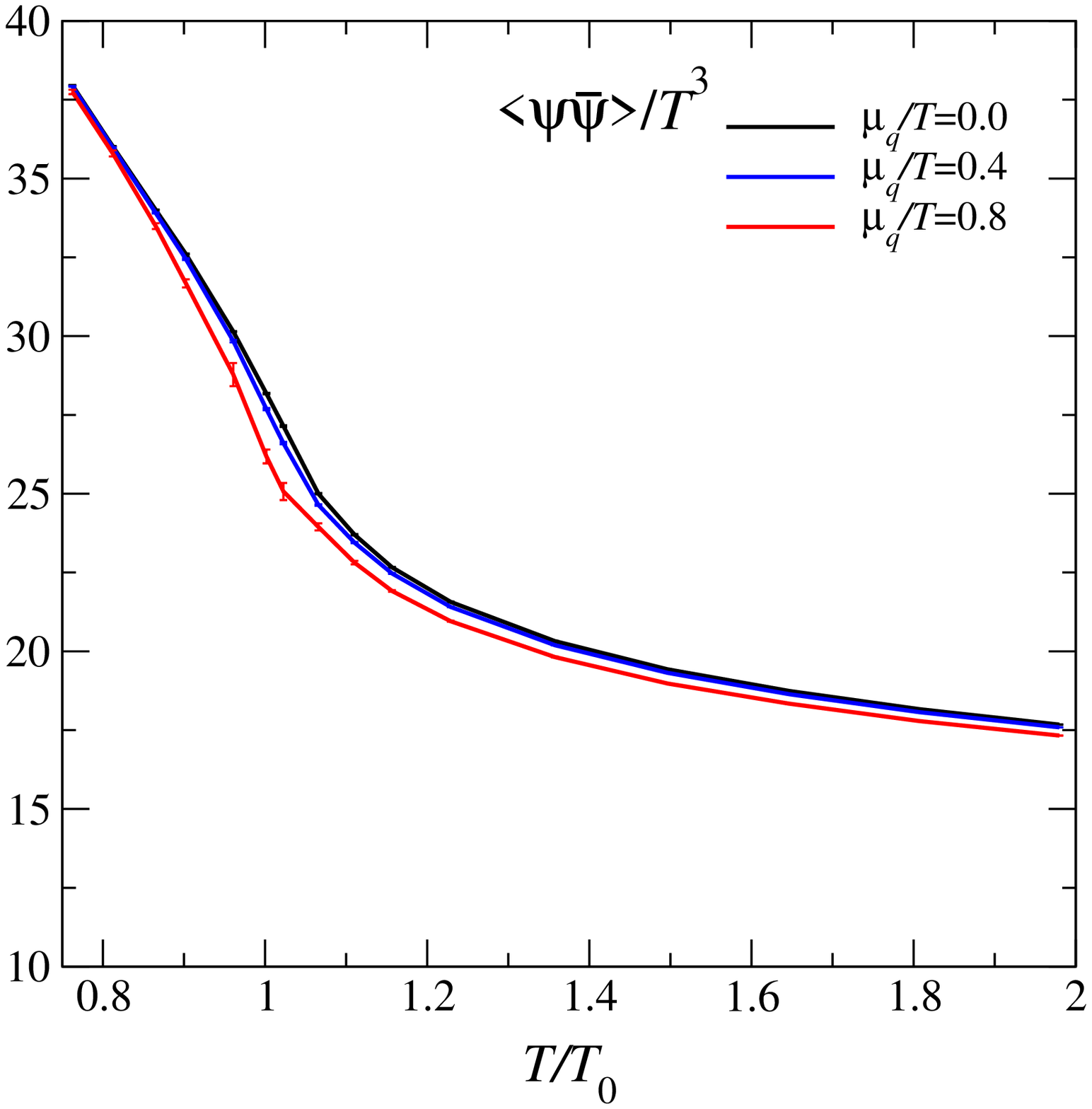, width=7.4cm}
\epsfig{file=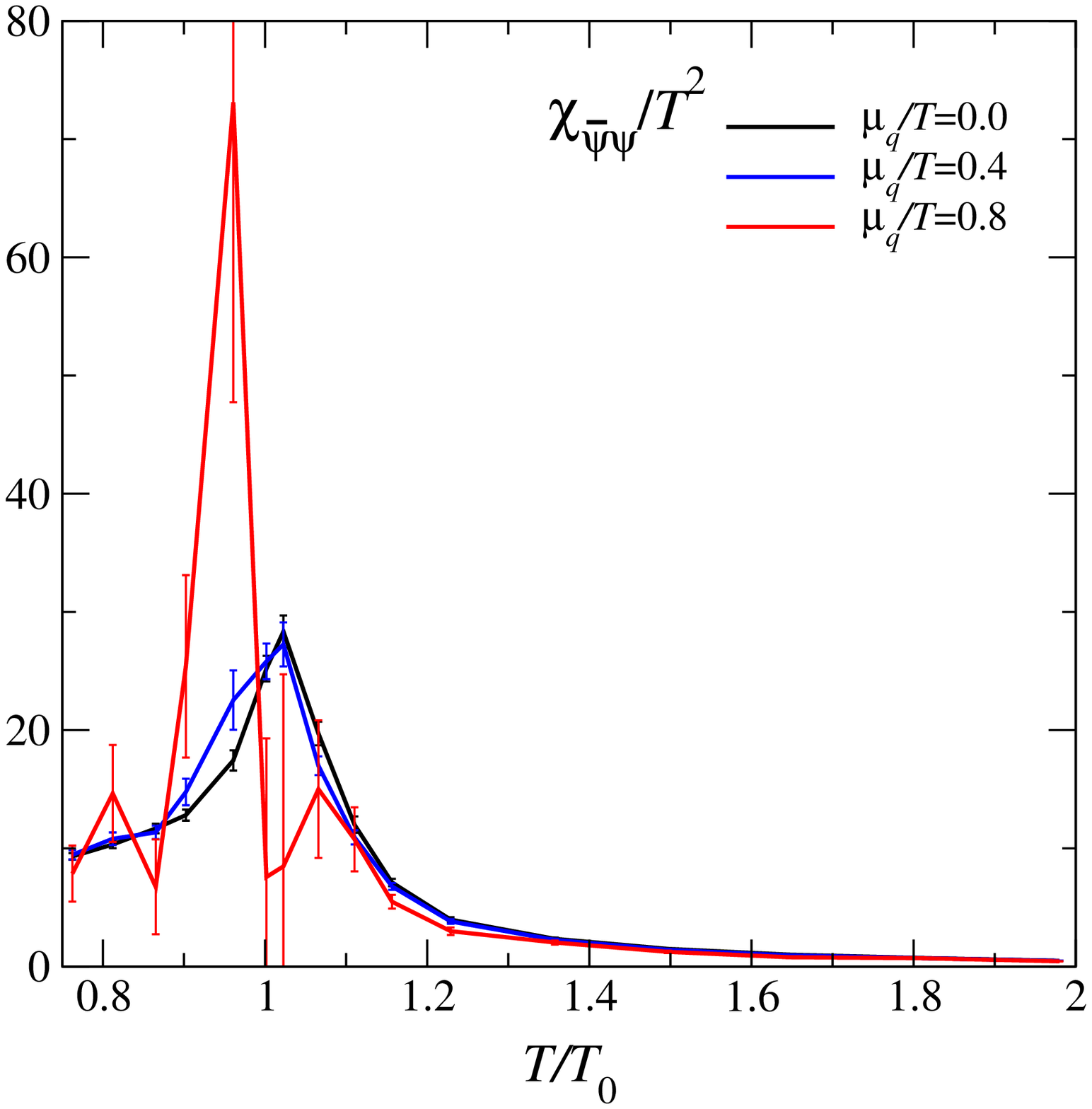, width=7.4cm}
\smallskip
\caption{The chiral condensate $\langle \bar{\psi} \psi \rangle$ (left)
and chiral susceptibility $\chi_{\bar{\psi} \psi}$ (right)
as a function of $T/T_0$ for $\mu_q/T=0,~0.4$ and 0.8. The chiral
condensate drops with increasing $\mu_q/T$ and the peak in
$\chi_{\bar{\psi} \psi}$ becomes more pronounced.}
\label{fig:ccn}
\end{center}
\end{figure}

Summarizing this subsection, the results of numerical simulations
presented in
\cite{BS_mu6} does not found the critical point. Instead, they
show a dramatic change of the baryon contributions
to thermodynamics, from a resonance gas with unmodified
masses, to a completely different regime in QGP. Chiral accessibility
possibly show reduction in the sigma mass by a factor of 2
 for $\mu\approx .8 T_c\approx 130 \, MeV$, which however remains
rather uncertain. 

\section{Nuclear forces}
  Studies of the NN forces, are at the very basis of nuclear physics,
go back by many decades and are discussed
in textbooks in detail. We will not discuss spin-spin or spin-orbit
parts of it, but
focus rather on one model, as simple as
possible, for the central forces. A well known Walecka model \cite{SW} 
is sufficient to demonstrate our main points.

Those forces can be viewed as a combination of
the sigma exchange\footnote{Strong variation of the phase shift
at 400-600 MeV is known for a very long time
 in $\pi\pi$ scattering, 
and was also identified in the attractive part
of the nuclear forces, see  e.g. a review
on applications of Walecka model \cite{SW}. 
Whether one would like to call it a resonance or not,
the fact remains that it dominates the 
attractive part of NN
interaction, and is thus responsible for nuclear binding and for
 our very existence. 
}
and the omega exchange terms
\be \label{walecka}
V= -{g_s^2 \over 4\pi} {e^{-r m_\sigma} \over r}+{g_v^2\over 4\pi}
 {e^{-r m_\omega} \over r}\ee
In this simple model $g_s=M_n/f_\pi\approx 10$ and 
$g_V^2=190$ which fits the data well enough.

The most important lesson about nuclear forces we would like to
remind the reader is that the
nuclear potential  is in fact a {\em highly tuned small difference}
 of
two large terms. Although both terms are comparable to the nucleon
mass,
the resulting potential has a minimum of only about a percent of
the available mass, $V(r_{min})/2 M_N \sim O(10^{-2})$.
 \footnote{Of course, in 
 nuclei there is another cancellation between kinetic and potential
energy, leading to a typical nuclear physics scale to $1\, MeV\sim
10^{-3} M_N$.} The so called ``relativistic systematics'', the dependence of 
nuclear forces on relative motion, reveals this fact well, as the
scalar
and vector parts change differently under Lorentz transformation. 
It makes the forces repulsive at semi-relativistic
energies, as multiple NN and heavy ion low energy
collisions have well documented. 

Because of this fine tuning, the nuclear forces are quite sensitive
to the 4 parameters of the model, two couplings and two
masses\footnote{Strong
sensitivity of the deuteron binding to $m_\sigma$ and eventually
to the ratio of quark masses to $\Lambda_{QCD}$
, combined with the Big
Bang
Nucleosynthesis data, makes the best limits on the
cosmological variation of these quantities \cite{FS}.  }.

If one simply takes $m_\sigma \rightarrow 0$, keeping the other
parameters the same, one gets huge attraction $V\sim - 1 GeV$.
If on the other hand the omega mass is also goes to zero, the result
is huge repulsion\footnote{Because the omega coupling is larger than
  the sigma one.}
$V\sim + 1 GeV$ 
. More moderate version of the same exercise
(with parameters to be explained later) 
 is demonstrated in Fig.\ref{fig_Walecka_pots} .

\begin{figure}[h!]
\centering
\includegraphics[width=8cm]{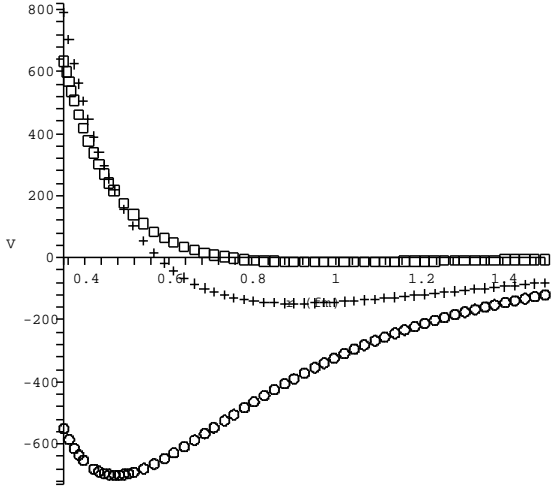}
\caption{\label{fig_Walecka_pots}
The NN potentials in the Walecka model, for the following values of
the masses:
$m_\sigma,m_\omega=$ 600,770 MeV (black), 280,770  (red), 280,500 (blue)
}
\end{figure}

A nucleon in a homogeneous matter is in a mean field potential
which can be easily calculated in this model.
The modification of it due to modification of
the $\sigma$-exchange diagram\footnote{What exactly is a sigma field
 and how it is interaction with pions and nucleons is 
described e.g. in \cite{CEG}}
\be \label{eq_Nmassshift}
\Delta V=- n_s\large({{g_s^*}^2 \over
  {m_\sigma^*}^2}-{g_s^2 \over m_\sigma^2} \large) \ee
where $n_s$ is the total scalar density of matter. Unlike vector
density, it includes all quarks and antiquark with the positive sign, so
\be n_s=n_B+n_{\bar B}+{2\over 3}n_{nGm} \ee
where nGm stands for all non-Goldstones mesons. The factor 2/3 corresponds to 
a simple additive approximation, assuming that the scalar coupling $g_s^2$
is additive for valence quarks.
\section{Vector mesons}

The mass shifts for  $\rho,\omega$ due to sigma exchanges
are of the same nature as for baryons, so in the same
spirit of additive coupling one expects extra\footnote{Which obviously
goes on top of other effects, such as resonances and the
sigma exchanges with unmodified mass.} mass shift to be
\be \label{eq_Rmassshift}
  \Delta V_{\rho,\omega}={2\over 3}\Delta V_N=-{2\over 3}n_s \large({{g_s^*}^2 \over
  {m_\sigma^*}^2}-{g_s^2 \over m_\sigma^2} \large) \ee

These shifts were experimentally seen in heavy ion 
collisions e.g. in peripheral AuAu collisions at RHIC, by STAR
collaboration \cite{STAR_Fachini} for few resonances, including
vectors $\rho$ and $K^*$. In the former case the mass shifts is about
$\Delta M_\rho=-70\, MeV$, with about half of that estimated to come
from
the sigma exchange \cite{BS_resonances}. If so, one may thing that
at the edge of the ``oval'' this term would be 4 times larger, or
about
$-120 \, MeV$. Very close to the critical point the effect is still
finite, in spite of zero sigma mass, due to vanishing coupling: this
however
only happens when sigma size gets really large.

The resonances like $\rho$ seen via $\pi\pi$ channel are of course
observed near the kinetic freezeout conditions, that is at rather
dilute matter. One may wander if those are not too far from the
critical line and critical point, to show any observable effect.

Another well known possibility
is observation of vector states via dileptons: in this case 
they are collected not from freezeout but from all the 4-volume of
expanding fireball, with the inside of the ``oval'' included.
Since we expect additional shift of vector mass, the enhancement
is supposed to increase roughly as $exp(\Delta M(T)/T)\sim 3$ 
due to twice lighter sigma.

We do indeed know that NA45 (CERES) collaboration have seen rather large
enhancement
of small mass dileptons at SPS, especially at 40 GeV. Recent run of
NA60
experiment with dimuons is expected to provide additional
clarification on this issue. A non-monotonous dependence of the 
light dilepton continuum on collision energy would be possible
indication for the critical point.

\section{The interaction of the pions }
\label{sec_pipi}
 Unlike baryons and non-
Goldstone modes of the theory, the pions 
cannot simply linearly interact with the sigma field, because any
nonzero scalar density would made them massive even in the chiral
limit and thus violate the Goldstone theorem. In the original
sigma model formulation that is resolved by some cancellation
between diagrams.   Better solution is to change the scalar field
to its ``radial'' version on the chiral circle (see \cite{CEG}),
after which sigma interacts only with a derivatives of the
pion field and preserves the theorem in each vertex.

  As in the chiral limit
 the interaction of low energy pions is described by Weinberg
  Lagrangian, 
\be L_W={1\over 2f_\pi^2} (\partial_\mu U^+)(\partial_\mu U) \ee
the corrections due to nonzero scalar density of the
medium can simply be absorbed into a modified coupling  
\be {1\over f_\pi^2}\rightarrow {1\over f_\pi^2}+n_s*G \ee
to be substituted into known results
about pion gas thermodynamics  \cite{Shu_80,Leutwyler}
and kinetics \cite{pionkinetics}.
For low-T pion gas the same is true for
 realistic QCD with massive quarks and pions, since 
 $O(m)$ corrections to scattering (such as Weinberg scattering
lengths) are very small.   

 A real hadronic matter produced in heavy ion collisions
have sufficiently high
temperature $T>T_f\sim 100 \, MeV$ to excite resonances, which truly
dominate the interaction.
 For pions those are $\sigma$ and $\rho$. those provide
scattering rate and ImV.  
The real part of the mean field potential for pions is very small
and usually neglected. 

The (momentum-dependent) potentials are usually defined as 
\be V(p)=(m^2+p^2+\Sigma(p))^{1/2}-(m^2+p^2)^{1/2} \ee 
where 
\be \Sigma_j(p)=\sum_i \int {d^3k \over (2\pi)^3} {n(\omega_k/T)\over 2
  \omega_k}M_{ij}(k,p)
 \ee is the mass operator related to the forward scattering
amplitude $M$. The sum stands for all particle types in matter.

 The amplitude can in turn be written as a sum over resonances
in each channel, and we are going to focus on $\sigma$ in $\pi\pi$.
For this case the forward scattering amplitude is of standard
Breit-Wigner type
\be  M_\sigma=-{4\pi E\over q}{\Gamma_{in} \over m_\sigma-E+i\Gamma_{tot}/2} 
\ee
where $E=\sqrt{s}$ and $q$ are the CM energy and momentum of two
colliding pions with momenta $\bf{p,k}$. We used the total width
in the form $\Gamma_{tot}=(.4 \, GeV)(q/E)$ where $q,E$ are the
C.M. momentum and energy per pion: not that it vanishes at q=0,
as is needed by the phase space. The $\Gamma_{in}=\Gamma_{tot}/2$.

As it is well known, the real part of the
amplitude {\em changes sign} at the resonance, and the sign means that
effective potential obtains positive
( $repulsive$) contribution
from $\pi\pi$ states above the resonance, $E>m_\sigma$, and
negative ($attractive$)  from the states below it. These two
parts of the integral over $k$ 
  tend to cancel each other. As a result,  the interaction between
 pions
is weakly attractive in a pion gas, with a potential strongly
dependent
on the pion momentum $p$.

\begin{figure}[t]
\centering
\includegraphics[width=8cm]{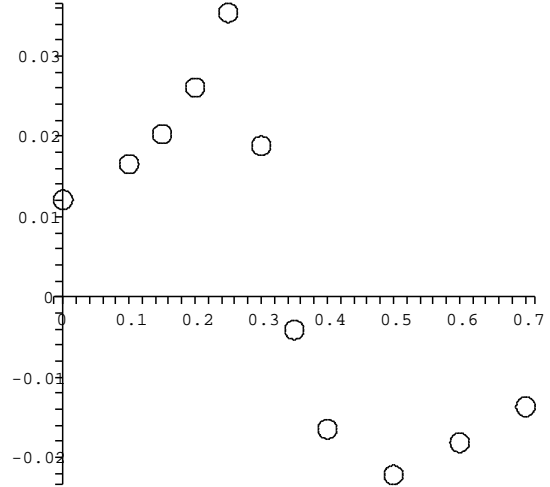}
\vskip 0.4cm
\caption{\label{fig_crit_pion_pot}
Effective potential for a pion at rest $Re[V_{eff}(p=0)]$ [GeV]
induced by sigma resonance, as a function of the sigma mass $M_\sigma$
[GeV].
}
\end{figure}

Another (somewhat more intuitive)) picture of pion in a pion gas
is that since they spend part of the time rotating around each other
in form of resonances,   
 the average pion's velocity 
 is slightly reduced in the pion gas relative to that in the vacuum
  \cite{pionliquid}.

However this is going to change near the QCD critical point. Near
the dashed line in Fig.
\ref{fig_crit_point_ovals}  the effective pion potential rapidly changes
and is getting repulsive  inside the solid shaded
oval, as seen from Fig.\ref{fig_crit_pion_pot}. 
For light enough
sigma, 
 all colliding pions are always $above$ the sigma resonance,
and thus the attractive contribution is absent, with only repulsive
contribution of sigma resonance remaining. As seen from the figure,
it leads to a rapid change of the potential, which
is not small. It happens exactly at
 the boundary of the oval, where $m_\sigma=2m_\pi$ and  
the scattering length of two pions at rest  goes to infinity
since the amplitude is $\sim 1/q$ and the relative momentum
$q\rightarrow 0$ here. The corresponding $\pi\pi$ 
 cross section at this point is reaching its s-wave unitary limit
\be \sigma\approx {4\pi\over q^2}\ee
 This behavior is identical to the so called
``Feshbach resonance'' at zero energy,
which is used in atomic physics for cold trapped atoms. The result is known to
be a  new ``strong coupling'' regime of matter in which it shows
liquid-like behavior \cite{atomic_flows}.

In summary of this subsection: we predict that close to
QCD critical point the pion-pion interactions rapidly change
their sign, leading to
repulsive mean potential.

\begin{figure}[t]
\centering
\includegraphics[width=9cm]{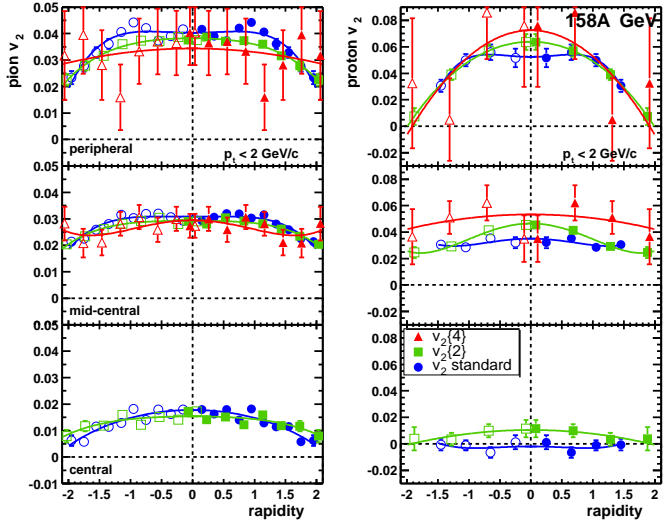}
\vskip 0.4cm
\caption{\label{fig_pb158}
Elliptic and directed flows of pions and protons
 versus rapidity at 158 
A$\cdot$GeV Pb+Pb collisions \protect\cite{NA49} measured for
three centrality bins:  central (dots), mid-central (squares) and
peripheral (triangles).  The solid lines are polynomial fits to the
data \protect\cite{NA49}.
}
\end{figure}

\section{Directed flows}
One general difference in  the vicinity of the critical point
is that near it the hydrodynamical description is to be appended
by simultaneous solution for the critical slow mode, the $\sigma$
field, coupled to matter
evolution\footnote{This is similar to what is done e.g. for
  magnetohydrodynamics, where one component
of all fields is singled out from the rest of the matter, and its
evolution
is treated separately, coupled to the stress tensor and hydrodynamics.
}. The first pioneering attempt to do so has
been done recently by Dumitru, Paech and Stocker \cite{DPS}.

We have argued above that near the critical point
the nucleon-nucleon and nucleon-non-Goldstone-meson interaction
should gets much more attractive than it is in vacuum. We will now
argue that it should affect collective flows of nucleons, reducing
their radial and elliptic flows. At the same time, we have argued 
in the previous chapter that the
pion-pion interaction gets more repulsive, so we expect the effect
of the opposite sign in the pion flows. 

Potentials related to collective sigma field add to 
pressure effect in defining the flows. 
$\nabla p$ term gets addition, $n_s \nabla \sigma$.
The distribution of the mean field sigma is to be found
from the distribution of the scalar density via
\be (-\nabla^2+m_\sigma^2)\sigma(x)= n_s\ee
(footnote: why not time derivatives? expansion is still slow)
which should be solved together with hydro equations.

 Treating its effect perturbatively
\be {\delta v \over v} \approx {\int dt n_s \nabla \sigma\over
 \int dt \nabla p} \ee
where the integrals are over the world line of each matter cell.
 
Let us make a very simple order-of-magnitude estimate for the
additional contribution to the pion flow we expect. The oval of 
fig.\ref{fig_crit_point_ovals} projected to a fireball will correspond
to also an ellipsoid-like
surface. The pions crossing from inside out would experience
a kick from the change in potential, repulsive inside and attractive
outside. Non-relativistically, an extra velocity this kick will produce is
\be \Delta v_\pi\approx ({2 \Delta V_{eff} \over m_{eff}})^{1/2}\approx .5\ee
where the final value in
the r.h.s. is obtained by using the change in the potential $\Delta
V_{eff}=40\, MeV$ from Fig.\ref{fig_crit_pion_pot}
and the effective pion mass for
motion in one direction
$m_{eff}^2=<p_t^2>+m_\pi^2\approx (2T_c)^2 $. 
This should be compared to collective radial velocity of matter
at SPS, which also happen to be
of the same magnitude, $<v_r>\sim .5$: the conclusion is the two effects
are comparable. Thus expected the non-monotonous  change
in pion radial flow  can be quite noticeable. It will be also
accompanied in increased elliptic flow of comparable magnitude.

Similar estimate can be made for the nucleon would include
the potential change\footnote{Note that the usual nuclear potentials
would not be enough.} $\Delta U\approx .2 GeV$ from
Fig.\ref{fig_Walecka_pots}
and $m_{eff}^2=<p_t^2>+m_N^2\approx 1\, GeV$,  resulting
in
\be \Delta v_N\approx -.5-.6 \ee 
Again, this is comparable to overall matter flow: but since
the sign is now $negative$ the total effect  expected is
cancellation
of flows. Needless to say, any uncertainties in  $\Delta U$ and schematic
estimate like that cannot substitute for a detailed calculation:
we how conclude that there is at least a potential for
{\em dramatic decrease} of both radial and elliptic nucleon
flows.

\subsection{ Flows observed at SPS}
 Let us now compare these ideas with observations. 
As it is well known, elliptic flow changes sign inside the AGS
energy domain and is quite well seen at its highest
energy, 11 GeV/N. The nucleon and pion elliptic flows 
are also rather large at the highest SPS energy 158 GeV/N,
and about double at RHIC.
 
The summary of the 158 GeV/N elliptic flow data from NA49 experiment
is shown in Fig.\ref{fig_pb158}. The magnitude is enhanced at
mid-rapidity, both for pions and nucleons.
As is the case in other cases, and the pion and nucleon $v_2$ are
comparable in magnitude.

\begin{figure}[t]
\centering
\includegraphics[width=9cm]{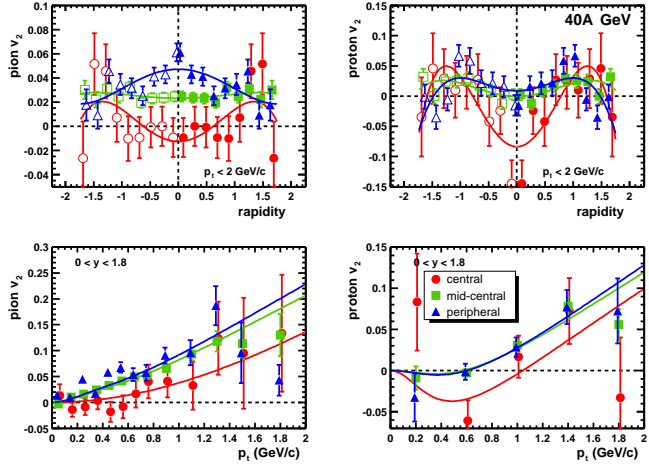}
\vskip 0.4cm
\caption{\label{fig_pb40}
Elliptic flow $v_2$ of protons versus rapidity from 40
A$\cdot$GeV Pb+Pb collisions \protect\cite{NA49} measured for
three centrality bins:  central (dots), mid-central (squares) and
peripheral (triangles).  The solid lines are polynomial fits to the
data \protect\cite{NA49}.
}
\end{figure}

  It is however not so at 40 GeV/N, as 
 the same NA49 experiment
finds 
  Fig.\ref{fig_pb40}. Such collapse of flow
at mid-rapidity is seen for all centralities and methods.
It is furthermore accompanied by a collapse of directed flow $v_1$:
this coincidence tells us that it is unlikely to be an experimental
problem and is probably real. 

Bratkovskaya  et al. \cite{Stocker} suggested that this unusual
behavior of the N flow may be a long-expected
manifestation of the ``softest point''. However that contradicts to 
the fact that
 elliptic flow of pions does not show
any collapse at this energy.

Our current proposal is instead that flows are
affected by the approaching critical point.
The nucleon flow collapse may be due to attraction effect we
discussed, while pions show the opposite potential. 

Needless to say, this  work is only exploratory in nature,
and its suggestion 
 should be investigated quantitatively in future work.
In particular, to study flows, one should
 not only include space-time-dependent sigma field and
potentials discussed above, but also
cascades which include continuous reactions like $N+\pi -> \Delta ->N+\pi
 $. Another issue worth considering is the effect of the critical mode
on strangeness flows,  by $\phi,\Lambda,\Xi,\Omega$. 

Finally, in view of all these ideas and observations, from theory,
lattice and even hints from data, it seems justified to revisit
the low SPS energy region. It can either be done with old SPS
detectors (NA49,NA60) in new dedicated runs, or at decelerated RHIC
beams,
or maybe in future GSI dedicated facility to be built in the next decade.

\begin{acknowledgments}
I am grateful to A.von Humboldt foundation for the generous award, supporting
my research on sabbatical leave in Germany.
This work is also partially supported by the US-DOE grants DE-FG02-88ER40388
and DE-FG03-97ER4014. I am grateful to C.Redlich for very useful discussion of
his unpublished work.
\end{acknowledgments}

\end{narrowtext}
\end{document}